\documentclass[12pt]{iopart}

\usepackage{graphicx}
\usepackage{amssymb}
\usepackage{color}

\begin{document}

\title{Surface defects and temperature on atomic friction}

\author{O~Y~Fajardo and J~J~Mazo}
\address{Departamento de F\'{\i}sica de la Materia Condensada and
  Instituto de Ciencia de Materiales de Arag\'{o}n, CSIC-Universidad
  de Zaragoza, 50009 Zaragoza, Spain}
\ead{yovany@unizar.es}

\date{\today}

\begin{abstract}
We present a theoretical study of the effect of surface defects on
atomic friction in the stick-slip dynamical regime of a minimalistic model.
 We focus on how the
presence of defects and temperature change the average properties of
the system. We have identified two main mechanisms which modify the
mean friction force of the system when defects are considered. As
expected, defects change locally the potential profile and thus affect
the friction force. But the presence of defects also changes the
probability distribution function of the tip slip length and thus the
mean friction force. We corroborated both effects for different values
of temperature, external load, dragging velocity and damping. We show
also a comparison of the effects of surface defects and surface
disorder on the dynamics of the system.

(Some figures in this article are in colour only in the electronic version)
\end{abstract}

\pacs{62.20.Qp, 81.40.Pq}
\submitto{\JPCM}

\maketitle

\section{Introduction}
Understanding friction is an actual scientific and technological
problem ~\cite{BowdenOxford1954,SingerKluwer1992,PerssonKluwer1996}.
Friction is a complex phenomenon of fundamental interest in many
scientific areas that occur at all length
scales~\cite{GneccoSpringer2007,
  UrbakhNature2004,BormuthScience2009,UrbakhNature2010}. Its
comprehension at the nanoscale is fundamental for instance for the
manipulation of nanoparticles and the miniaturization of moving
devices as nano-electromechanical systems (NEMs).  With the
development of experimental techniques such as the force friction
microscope (FFM) and the surface force apparatus (SFA), experimental
and theoretical studies of friction at the atomic scale have received
an increasing interest in the last years.  Furthermore, as the
frictional interface between two surfaces involves complex
interactions among many asperities, the atomic force microscope is an
exceptional tool to better understand friction at nanoscale level
since it can be described essentially as a single asperity dragged
along a surface~\cite{GneccoSpringer2007, UrbakhNature2004,
  SzlufarskaJPhysD2008}.

So far, most theoretical studies describing force friction microscopy
(FFM) experiments have focused on the behavior of defect{-}free and
perfect periodic surfaces with or without the inclusion of thermal
effects~\cite{JohnsonTribolLett1998,GneccoPRB2000,SangPRL2001,MedyanikPRL2006}.
However, the study of the effect of substrate disorder or defects on
atomic friction is particularly important since atomically flat
surfaces represent ideal models and defects of different kind are
always present. A recent study in a one-dimensional model indicates
that small uncertainties in the interaction effective potential
between the FFM tip and the surface can produce strong changes in the
frictional behavior of the tip~\cite{OFajardoPRB2010}. Other results
shows that other kind of imperfections in the substrate potential also
modify the frictional behavior at atomic scale~\cite{TshiprutPRL2009,
  BraimanPRE,ReguzzoniPNAS2009}. Reguzzoni {\it et al} studied friction in
the sliding of a xenon monolayer on a copper
substrate~\cite{ReguzzoniPNAS2009}. They found that the onset of slip
of the monolayer is strongly affected by the presence of vacancy-type
defects within the monolayer. H\"{o}lscher et al analyzed the load
dependence of atomic friction at atomic-scale surface
steps~\cite{HolscherPRL2008}.

In order to better characterize atomic-scale friction under
realistic surface potentials we present here results for the effect of
{\em surface defects} on atomic friction and its interplay with
thermal effects. By surface defects we refer to absorbed molecules,
vacancies or inclusion of attractive or repulsive atoms into the
perfect lattice. We will use a minimalistic model which focuses 
in a small number of the most relevant degrees of freedom and emphasizes 
the nonlinear nature of frictional dynamics.
We consider the one-dimensional case and focus on the
stick-slip region of the friction force versus dragging velocity
curve. The same problem has been studied previously by Tshiprut {\it et
al}~\cite{TshiprutPRL2009}. We present results for the mean value of
the friction force and the slip length and for the slip length
probability distribution functions (PDFs) for a range of values of the
corrugation potential amplitude, density of defects, temperature, and
damping.  We present results for four different types of defects. Our
results indicate that the presence of defects may strongly modify the
frictional behavior at atomic scale. The observed changes in the
friction force result from local changes of the potential profile,
which in many cases produce also significant changes in the PDFs of
the slip lengths.  We have compared our results with the defect-free
case and evaluated how the inclusion of defects locally modify the
slip length that the tip performs. To finish we make a detailed
comparison of the results obtained for {\it surface disorder} and {\it
  surface defects}. We also observe that the effect of surface
disorder on averaged quantities is screened at enough strong thermal
fluctuations. On the contrary, for the surface defects problem we
found significant effects even at high temperatures.

\section{Model}
We study a generalized Prandtl-Tomlinson model which includes the
thermal effects\cite{SangPRL2001},
\begin{equation}
\eqalign{ M\frac{\rmd ^{2}x}{\rmd t^{2}}+M\gamma \frac{\rmd x}{\rmd t}+\frac{\partial U(R,x)}{\partial x} = \xi(t), \cr
U(R,x)=\frac{k}{2}\left[ R(t)-x \right]^{2}+V(x).}\label{eq2} 
\end{equation}
	
Here the tip is modeled as a single particle dragged by an elastic
spring over a one-dimensional substrate potential. $U(R, x)$ accounts
for the tip effective potential and it includes the elastic coupling of
the tip with a support which moves at constant velocity $v_s$
($R(t)=R_0+v_s t$), and the tip{-}surface interaction $V(x)$.  $M$ and
$x$ are the effective mass and the lateral position of the tip and $k$
is an effective spring constant. $\xi (t)$ is the random noise term
which satisfies the fluctuation{-}dissipation relation $\langle \xi
(t) \xi (t ') \rangle = 2M \gamma k_{\rm B}T \delta (t-t') $ with $
\gamma $ the microscopic friction coefficient and $ k_{\rm B} $ the
Boltzmann constant.

\begin{figure}[!t]
\centering
\includegraphics[scale=0.25]{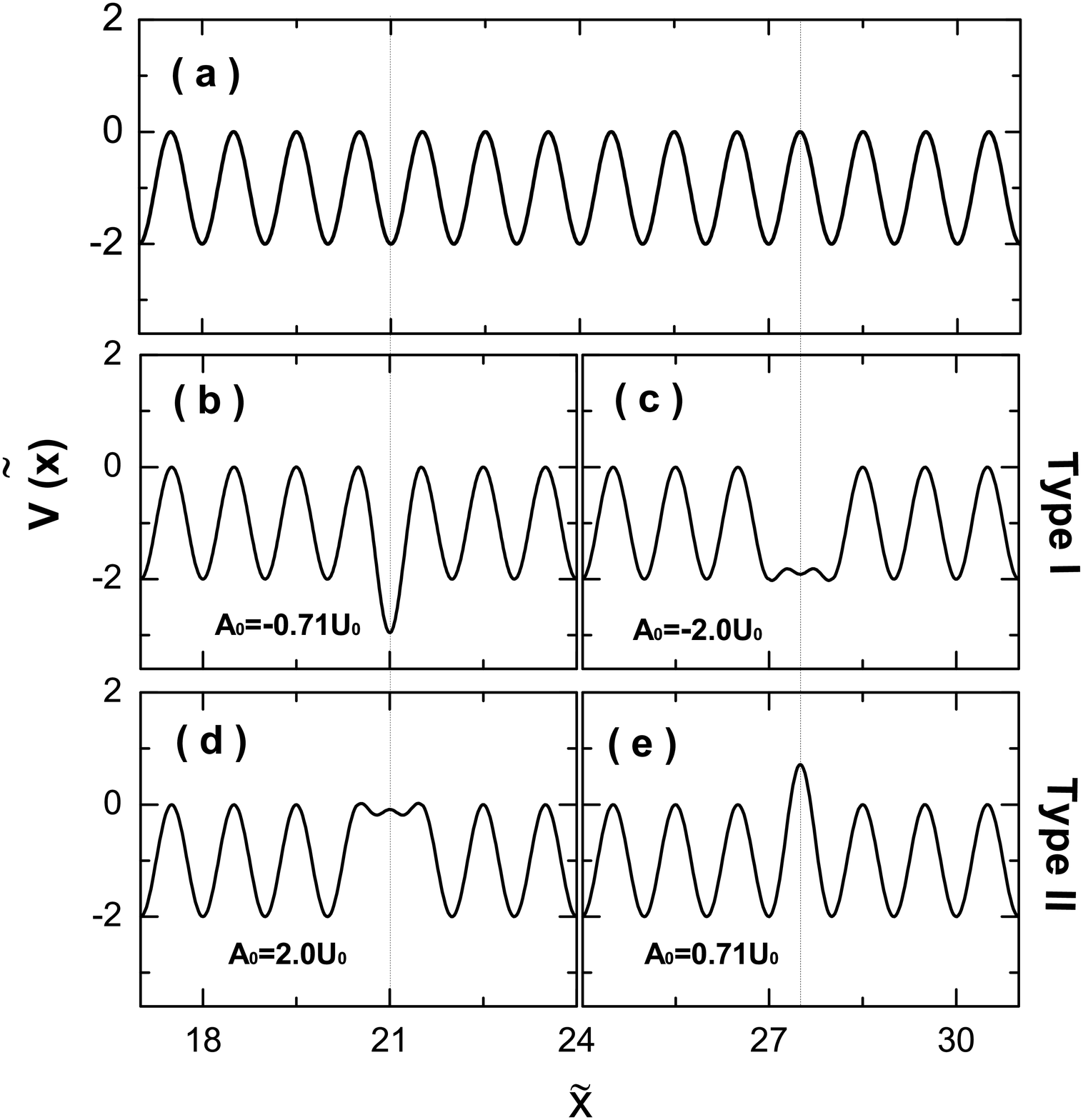}
\caption{Tip-surface potential $V(\widetilde{x})/U_0$ for (a) the perfect lattice,
  (b) the type I defect-in-minima lattice, (c) the type I defect-in-maxima lattice,
  (d) the type II defect-in-minima lattice and 
  (e) the type II defect-in-maxima lattice.}
\label{Fig1}
\end{figure} 

\begin{figure}[t]
\centering
\includegraphics[scale=0.31]{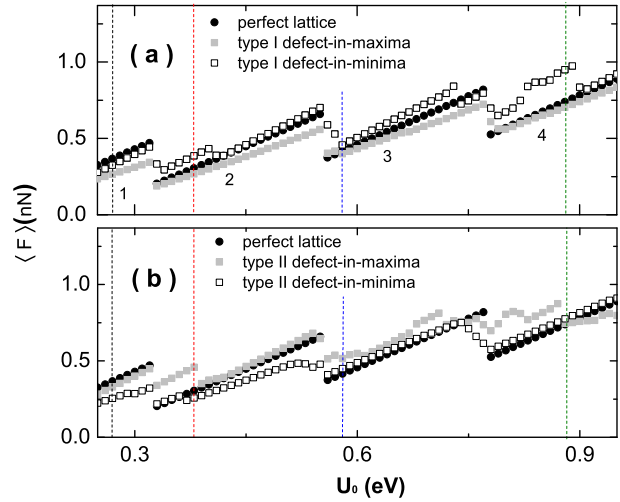}
\caption{ Average friction force versus corrugation
  potential amplitude $U_0$ in the low driving velocity region
  ($v_s=10$ nm/s). $T=0$ and $\gamma=10^5$s$^{-1}$. (a) results for
  the defect-free (solid circles), type I defects-in-maxima (solid
  squares), and type I defects-in-minima (open squares)lattices. (b)
  results for the defect-free (solid circles), type II
  defects-in-maxima (open squares), and type II defects-in-minima
  (solid squares) lattices.  The density of defects is $d=30\%$.}
\label{Fig2}
\end{figure}

\begin{figure}[h]
\centering
\includegraphics[scale=0.31]{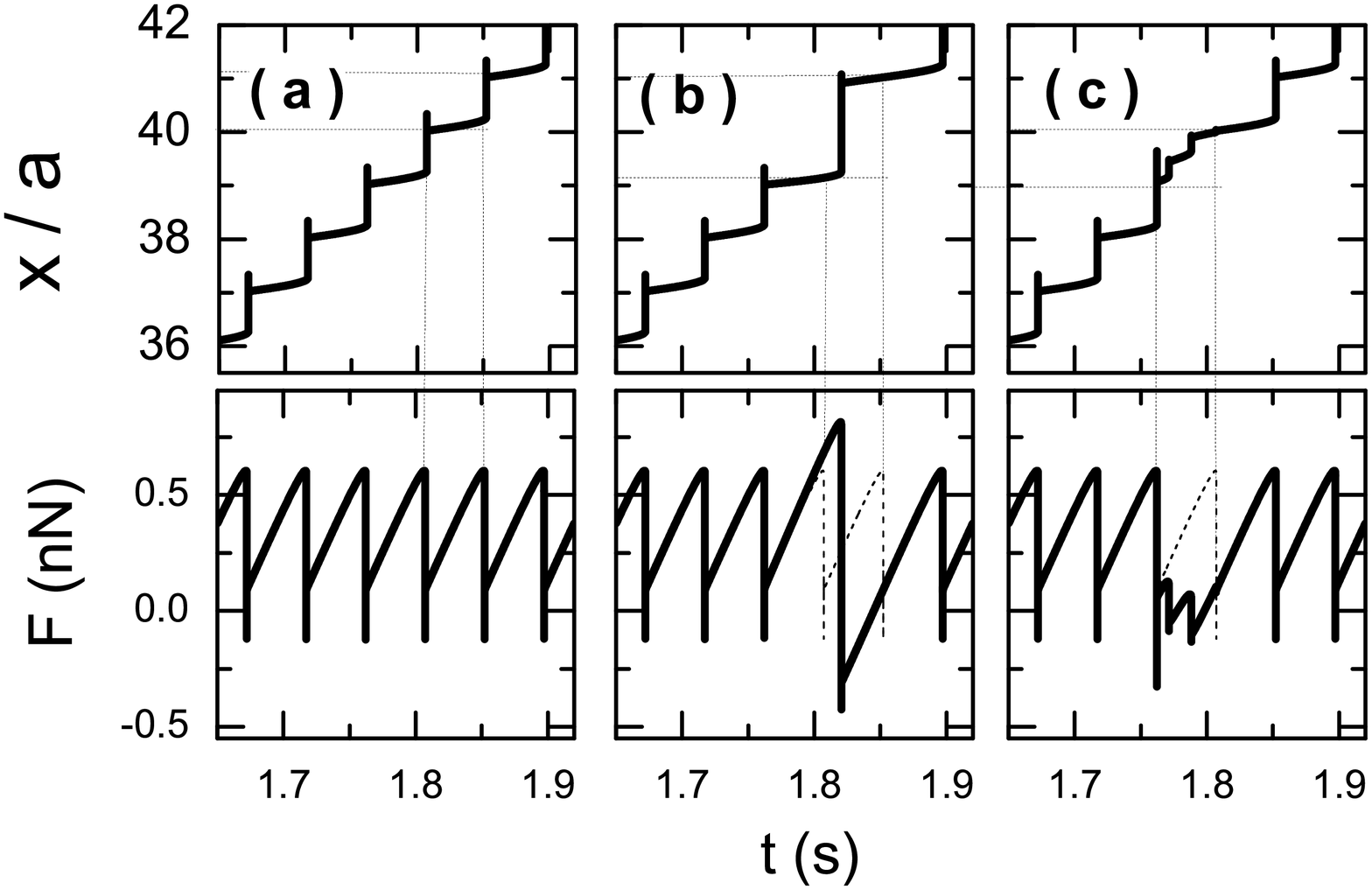}
\caption{ Figures (a) to (c) show the local effects of
  the inclusion of defects on the instantaneous position and friction
  force of the tip for $U_0=0.27$ eV. The density of defects is $d=30\%$,
  $T=0$ and $\gamma=10^5$s$^{-1}$. (a) stands for the perfect lattice,
  (b) for type I in minima and (c) for type I in maxima.}
\label{Fig3}
\end{figure}

We model surface defects by including gaussian terms in the standard
tip{-}surface interaction potential,
\begin{equation}
 V(x)=-U_0\left[ 1.0+\cos \left( \frac{2\pi}{a} x\right) \right] +\sum_{j}A_0\rme^{-\frac{(x-x_j)^2}{2\sigma^2}}.
\label{defect}
\end{equation} 	
$A_0$ gives the amplitude of the defect potential and $\sigma$ its
range. $a$ and $U_0$ are the lattice spacing and the amplitude of the
defect-free surface potential, respectively. In theory, this amplitude
can be changed by varying the normal
load~\cite{MedyanikPRL2006,SocoliucPRL2004}.

We will show below results for the four kind of defects shown in
figures~\ref{Fig1}(b)-(e). Panel (a) shows the potential profile for a
perfect lattice. As in~\cite{TshiprutPRL2009} we model the inclusion
of atoms of different nature by introducing a random series of
gaussian terms located in the minima ($x_{j}=na$) of the otherwise
perfect lattice (see figure~\ref{Fig1}(b) where $A_0=-0.71 U_0$ and
$\sigma=0.2 a$). The absence of atoms in the substrate is modeled
introducing gaussian terms located randomly in the maxima
[$x_j=(2n+1)a/2$] of the lattice (see figure~\ref{Fig1}(c) where
$A_0=-2.0 U_0$ and $\sigma=0.2 a$). Figure~\ref{Fig1}(d) stands for
defects located in the minima with $A_0=+2.0 U_0$ and $\sigma=0.2 a$;
and figure~\ref{Fig1}(e) for defects in the maxima with $A_0=+0.71 U_0$
and $\sigma=0.2 a$. For the first two kinds of defects,
figures~\ref{Fig1}(b)-(c), $A_0<0$ and we will refer to them as type I
defects. For the last two cases, figures~\ref{Fig1}(d)-(e), $A_0>0$ and
we will refer to them as type II defects.

\begin{figure}[!t]
\centering
\includegraphics[scale=0.7]{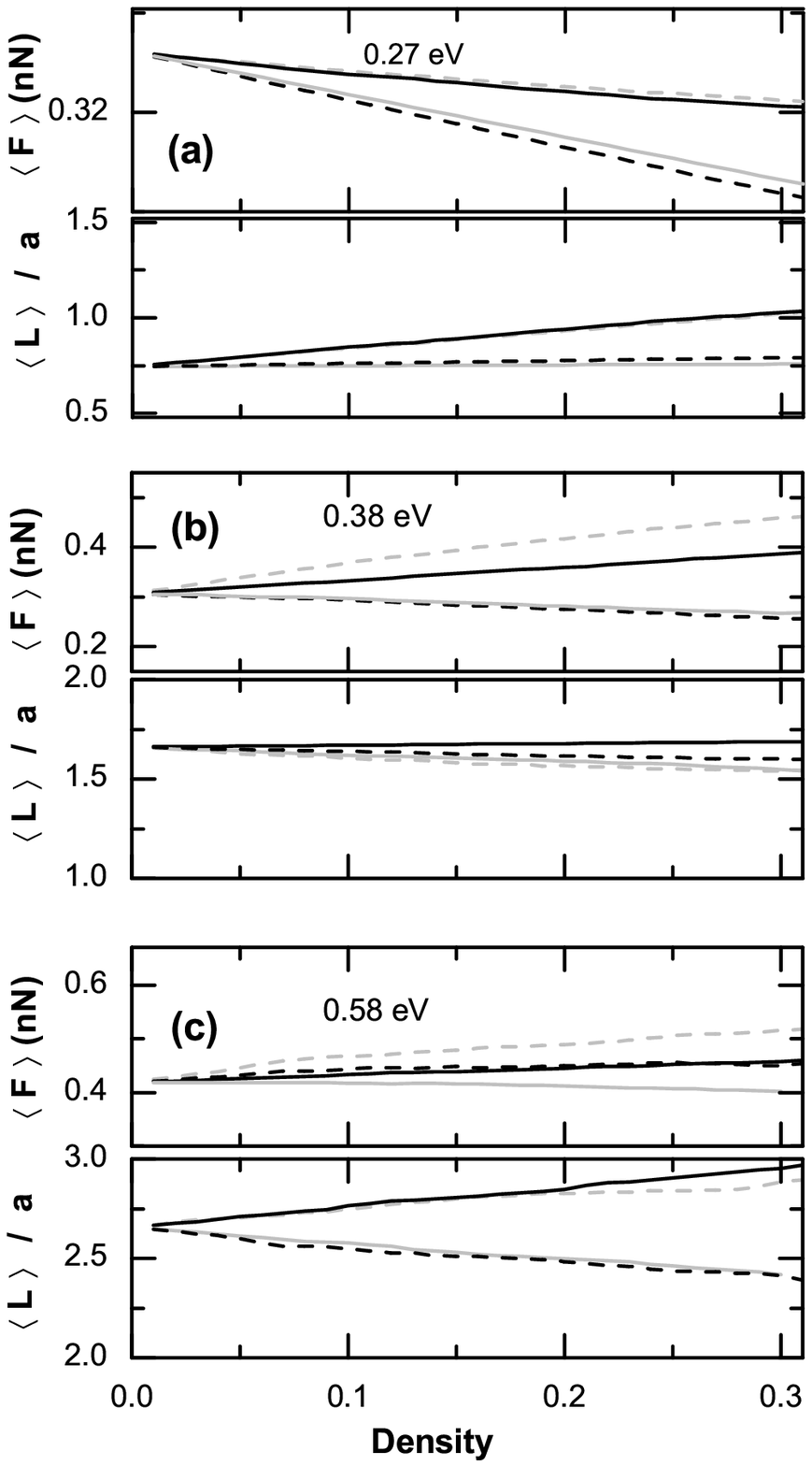}
\caption{ Average friction force and average length slip versus
  density of defects centered in minima (gray) or in maxima
  (black). Type I defects are continuous lines and 
  type II dashed ones. We show results for  $Uo= 0.27$ eV, $U_0=0.38$ eV, 
  $U_0=0.58$ eV. $T=0$ K and $v_s=10$nm/s.} 
\label{Fig4}
\end{figure}

\begin{figure}[tb]
\centering \includegraphics[scale=0.25]{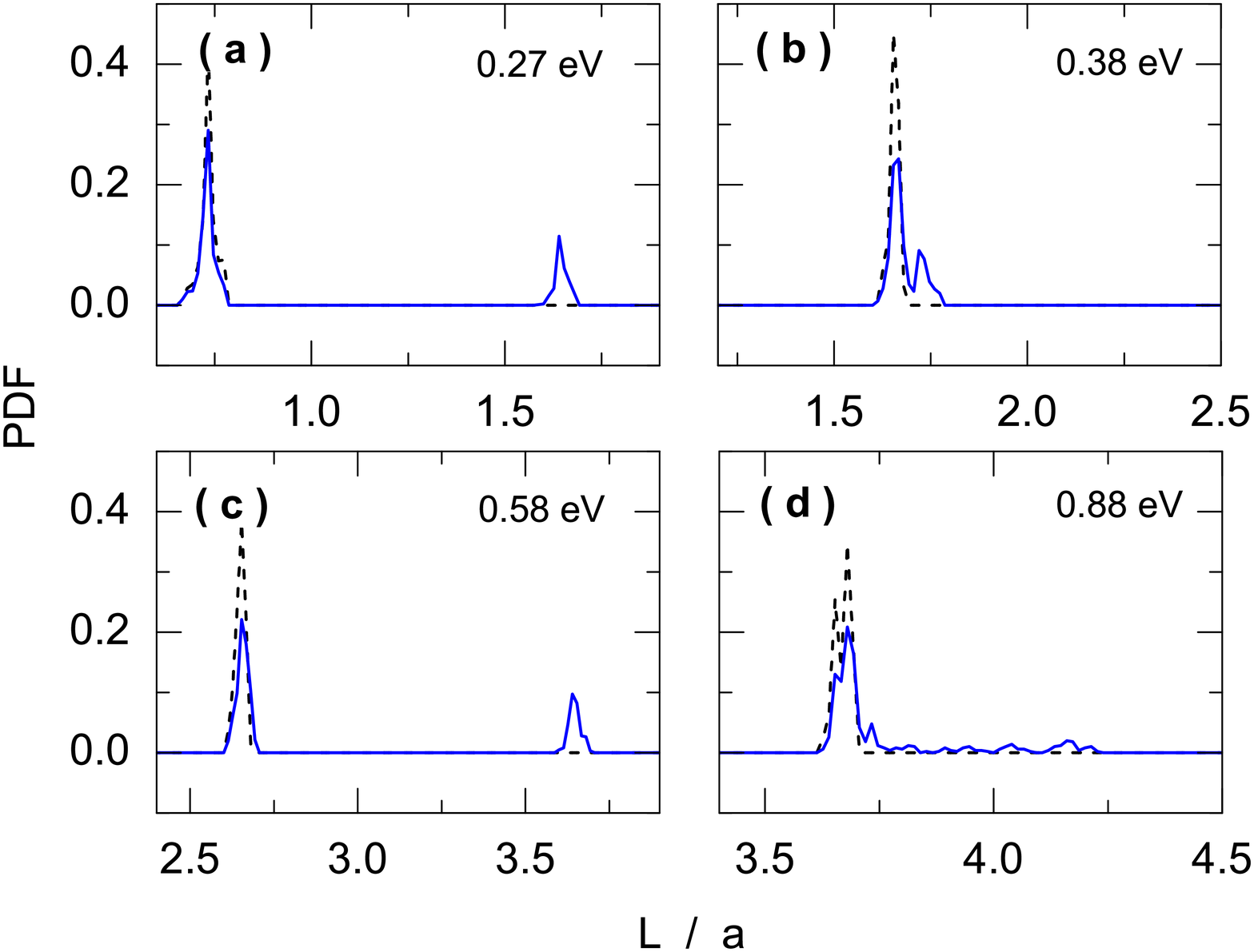}
\caption{ Probability density functions (PDF) of the
  slip length for the perfect (black dashed line) and the
  defect-in-minima (density 30\%, solid blue line) lattices at $T=0$ K
  and (a) $ Uo= 0.27$ eV, (b) $U_0=0.38$ eV, (c) $U_0=0.58$ eV and (d)
  $U_0=0.88$ eV.}
\label{Fig5}
\end{figure} 

To obtain dimensionless equations, energy can be measured in units of
the corrugation potential amplitude $U_0$, space in units of the
lattice spacing ($\widetilde{x}=2\pi x/a$) and time in units of the
natural frequency for oscillations of the tip in the surface potential
($\tau=\omega_p t$ with $\omega_p=2\pi\sqrt{U_0/Ma^2}$). Then
$\widetilde{\gamma}=\gamma / \omega_p$, $\widetilde{k}=1/\Theta=k a^2
/(4\pi^2U_0)$ and $\widetilde{v_s}=v_s\sqrt{M /U_0}$ are the
dimensionless damping, spring constant and velocity respectively. Some
of our results will be shown as a function of $U_0$. From the above
relations we see that a change in $U_0$ produces a change in both the
dimensionless damping and velocity of the system.

We have performed detailed numerical simulations of the dynamics of
the system at different parameter values.  We will present results for
the mean friction force $\langle F \rangle$, with $F(t)= k
[R(t)-x(t)]$, mean slip length $\langle L \rangle$ and slip length
PDFs.  Following the work by Tshiprut et al~\cite{TshiprutPRL2009},
we have used $M= 5.0\times 10^{-11}$Kg, $a=0.45$nm, $k=1.5$N/m,
$\gamma=10^{5}$s$^{-1}$ (except in section V where we change the
damping) and $U_0$ in the 0.2eV to 1.2eV range (then
$\widetilde{\gamma}$ goes from 0.3 to 0.1 and $\Theta$ from 4 to
25). For $v_s$ we have used 10nm/s,~\cite{Comentario1}. Our simulations 
were performed using an stochastic Runge-Kutta
method with dimensionless time step 0.01. Averages were computed after
a sliding distance of 2000 lattice constants.  An evolution of this
length allows to obtain results of the average quantities with
reliable statistics. We will show also the Probability Distribution
Function of the slip lengths.

\section{Surface defects without temperature}

Here we consider the effect of surface defects in the deterministic
dynamics of the system, $T=0$~K. We focus on the stick-slip region of
the force versus velocity characteristic curve. The well-known
behavior of the friction force as a function of the potential
amplitude for the defect-free
lattice~\cite{GneccoSpringer2007,UrbakhNature2004,SzlufarskaJPhysD2008,JohnsonTribolLett1998,GneccoPRB2000,SangPRL2001,MedyanikPRL2006,OFajardoPRB2010,TshiprutPRL2009}
is presented in figure~\ref{Fig2} (solid circles). There, the friction
force has been computed in the low driving velocity region ($v_s=10$
nm/s) with $U_0$ ranging from 0.25 to 1.2 eV.  A series of
discontinuities can be seen. They mark transitions between different
dynamical states characterized by a well defined mean value of the
length of the slip
events~\cite{JohnsonTribolLett1998,MedyanikPRL2006,OFajardoPRB2010,NakamuraPRB2005}.
Regions numbered as 1, 2, 3 and 4 in the figure correspond,
respectively, to regions characterized approximately by one, two,
three and four-lattice constant jumps. The value of the normalized
parameter $\Theta$ in the first four regions is: $ 1.0 < \Theta_{1}
\leq 6.66 \leq \Theta_{2} \leq 11.45 \leq \Theta_{3} \leq 16.03 \leq
\Theta_{4} \leq 20.61$.  For values of $\Theta < 1$, the tip
slides smoothly and stick-slip events do not appear.

Figure~\ref{Fig2} also presents the result for surface defects with
density $d=30\%$. As expected, in general the presence of type I
defects in the potential minima or type II defects in the potential
maxima increases the friction. On the contrary the suppression of
potential maxima, type II defects in potential minima or type I
defects in potential maxima, decreases friction. Thus, a system with
both types of defects will suffer a certain balance of both
effects. However, the dynamics of the system is much more complex and
this is not always the case. For type I in minima defects for
instance, it is easy to see the presence of two competing effects: a
deep potential causes a longer stick and then high friction, but it
can also cause longer slips, which reduce friction. Depending on the
parameter values the first or the second mechanism is the more
important one. Furthermore, close to the transition points surface
defect effects are difficult to predict. In addition, significant
changes are observed for large values of the potential $U_0$.  There,
the effective damping is smaller~\cite{Comentario2} and the dynamics
is more sensitive to small changes in the substrate potential.

\begin{figure}[tb]
\centering
\includegraphics[scale=0.31]{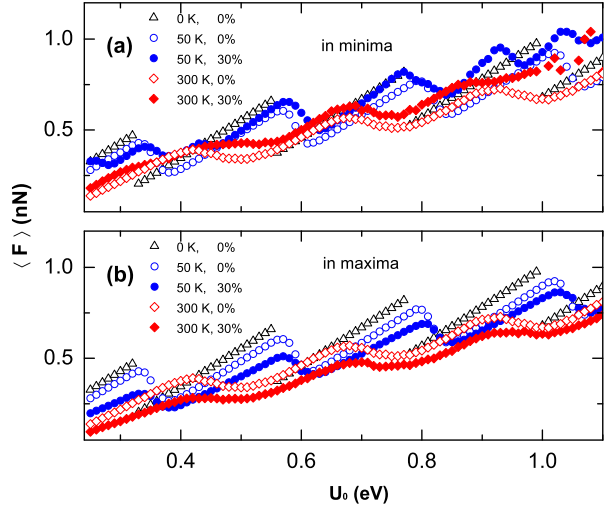}
\caption{ Average friction force versus corrugation
  potential amplitude $U_0$ at $T=0$, $50$ and $300$ K. (a) Type I
  defect-in-minima and (b) type I defect-in-maxima case. Open symbols
  are the defect-free lattice and solid ones for surface defect
  ones. In all the cases $v_s=10$nm/s.}
\label{Fig6}
\end{figure} 

\begin{figure*}[tb]
\centering
\includegraphics[scale=0.55]{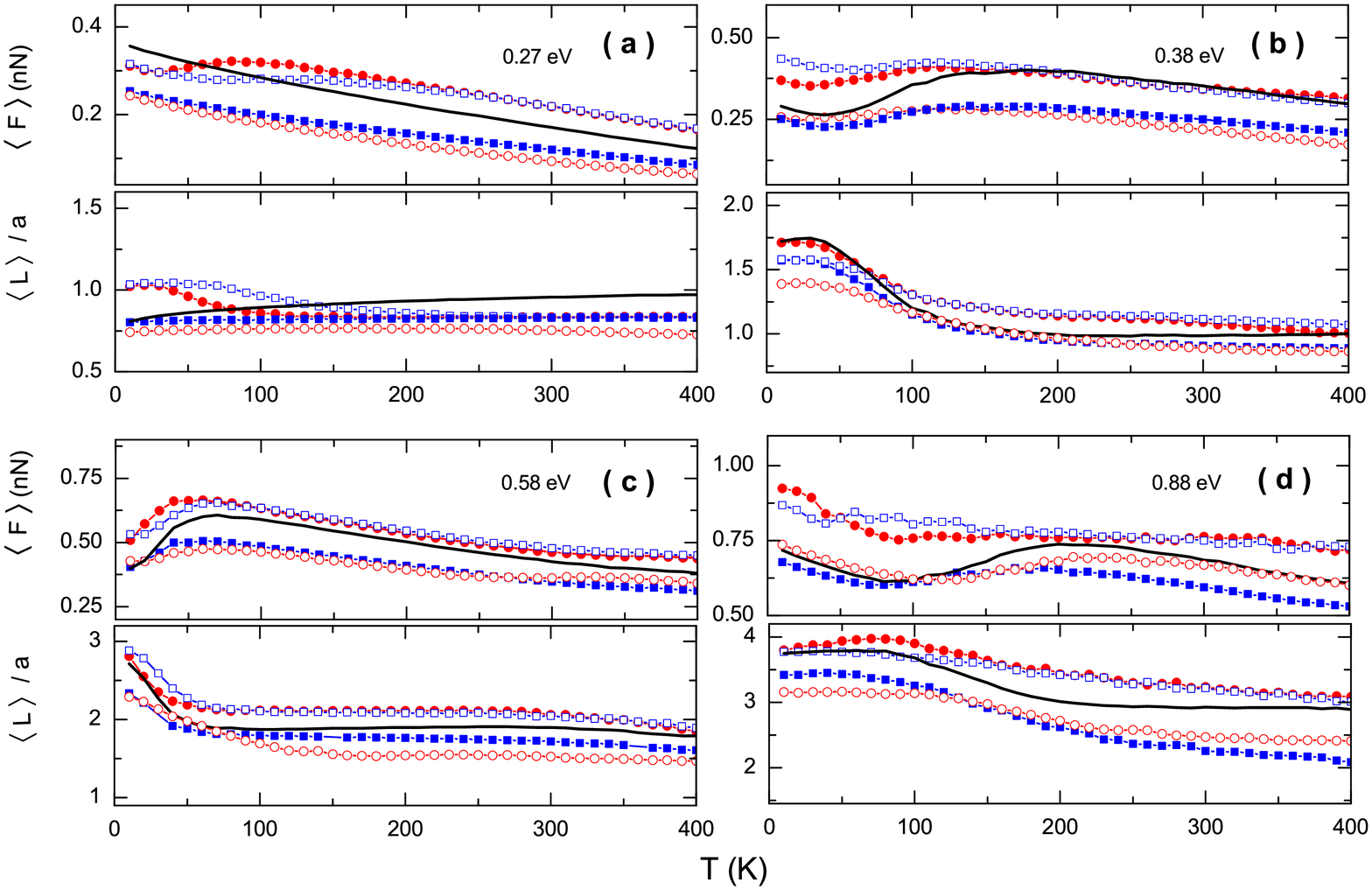}
\caption{ Average value of the friction force and the
  slip length versus temperature for different potential corrugation
  amplitudes $U_0$ for the defect-in-minima, defect-in-maxima and
  perfect (continuous black line) cases.  Type I defect-in-minima
  (solid circles), type I defect-in-maxima (solid squares), type II
  defect-in-minima (open circles) and type II defect-in-maxima (open
  squares). $U_0=$0.27, 0.38, 0.58 and 0.88 eV; panels (a) to (d)
  respectively.  As usually $v_s=10$nm/s.}
\label{Fig7}
\end{figure*} 

Figures~\ref{Fig3}(a)-(c) show the effect of the surface defects on
the instantaneous position of the tip (top) and friction force
(bottom). Figure~\ref{Fig3}(a) stands for the perfect sinusoidal
potential. This case is characterized by a regular tip dynamics with a
characteristic slip length and a periodic friction force.  The
inclusion of type I defects in potential minima, figure~\ref{Fig3}(b),
produces a significant change in the slip lengths and friction forces.
Finally, type I defects in maxima, figure~\ref{Fig3}(c), produce a
strong reduction of the slip length and a reduction of the stick
phase. We have checked that as expected such reduction is complete if
we totally remove the potential maximum.

Figures~\ref{Fig2} and~\ref{Fig3} were computed for a density of 30\%
of defects.  In figure~\ref{Fig4} we study the value of the mean
friction force and mean slip length as a function of the density of
defects (solid lines are for type I defects and dashed lines for type
II ones) for three values of the potential amplitude: $U_0=0.27$,
$0.38$ and $0.58$eV (marked in figure~\ref{Fig2}). We observe a
significant modification of the friction force. As it can be seen in
the figure, this change sometimes is associated to an important change
of the mean slip length, and other times is due to the modification of
the potential profile without change of $\langle L \rangle$. Figure
also shows that the friction force and the slip length change linearly
with the density of defects. Due to computational reasons, in what
follows we will show results for a density of defects $d=30\%$.

In order to get a better insight of the findings showed in
figure~\ref{Fig4}, we have computed slip length PDFs at different values
of $U_0$, figure~\ref{Fig5} for $d=30\%$. We show results for the
defect-free and type I defect-in-minima cases, other type of defects
can be understood in a similar way. For $U_0=0.27$ eV
[figure~\ref{Fig4}(a)], for the defect free case a single peak with
$\langle L \rangle$ close to one is found~\cite{Comentario3} [black dashed line in
  figure~\ref{Fig5}(a)]. For the lattice with defects the slip length
PDF (blue line) presents significant modifications. Now we see a
second peak in the distribution, located near to $L/a=1.65$ and a
decrease of the original one located at $L/a=0.73$. Thus the average
slip length is increased and the friction force reduced. Larger
densities produce an additional increase of the second peak and
reduction of the first one.

For $U_0=0.38$ eV the mean slip length $L/a \simeq 1.65$ for all the
defect density studied, figure~\ref{Fig4}(b). The corresponding PDF are
also slightly modified as seen in figure~\ref{Fig5}(b). However, we
observe an important increase of the friction force which in this case
is ascribed to the increase in the potential barriers associated to
the defects.

In the third case, $U_0=0.58$ eV, figure~\ref{Fig4}(c), we observe both
effects: an important modification of the slip length PDF with an
increase of the slip length mean value and also an increase of the
friction force.  

Finally, the last case shown in Figure 5(d), corresponds to a small
normalized damping and there the dynamics of the tip is different.  At
low values of the normalized damping tip oscillations exist between
the accessible wells.  This effect introduces additional changes in
the friction force and the slip length.  These two combined effects,
defects and low normalized damping, produce a slightly modified slip
length but a significant change in friction force.

\begin{figure*}[tb]
\centering
\includegraphics[scale=0.25]{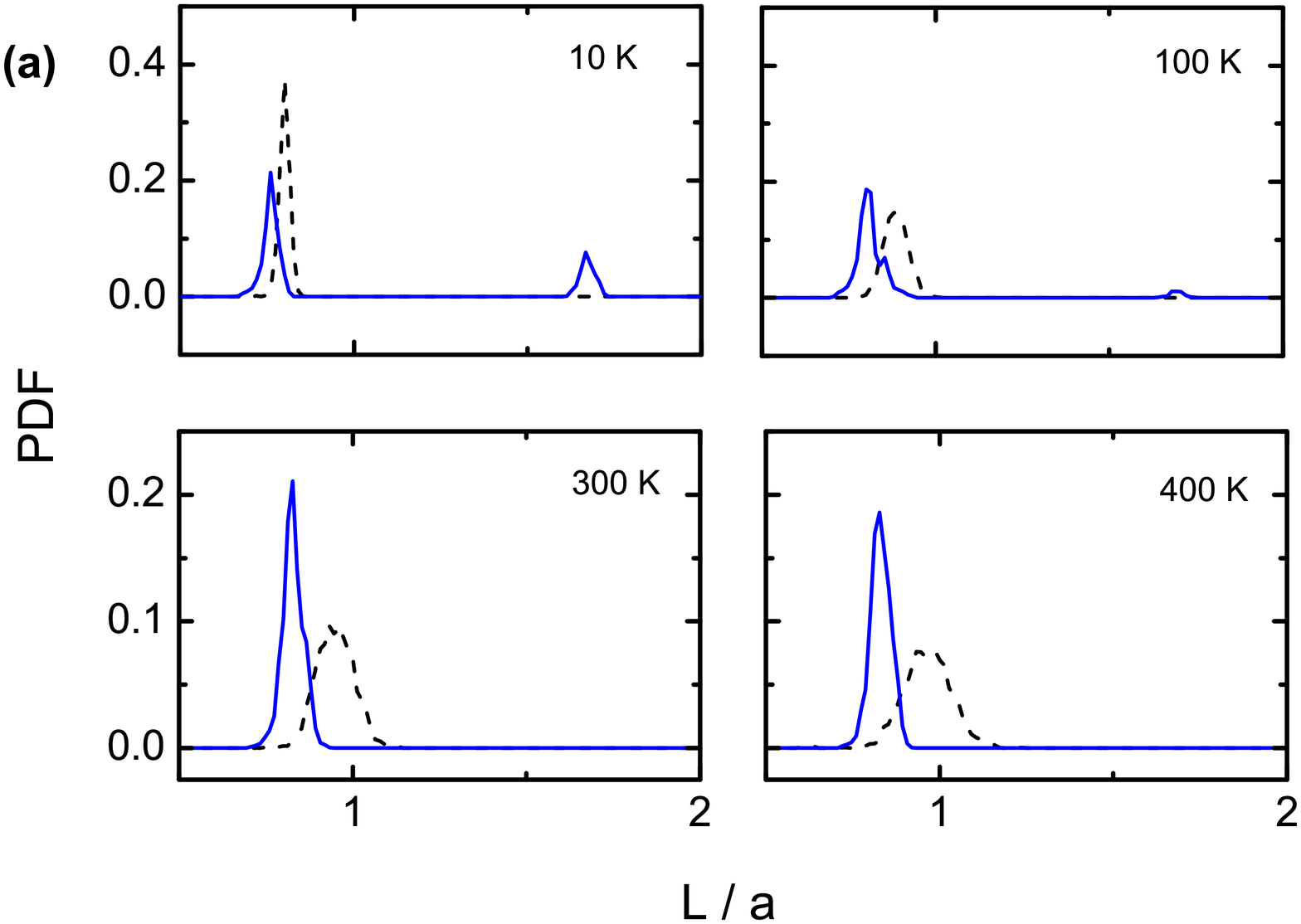}
\includegraphics[scale=0.25]{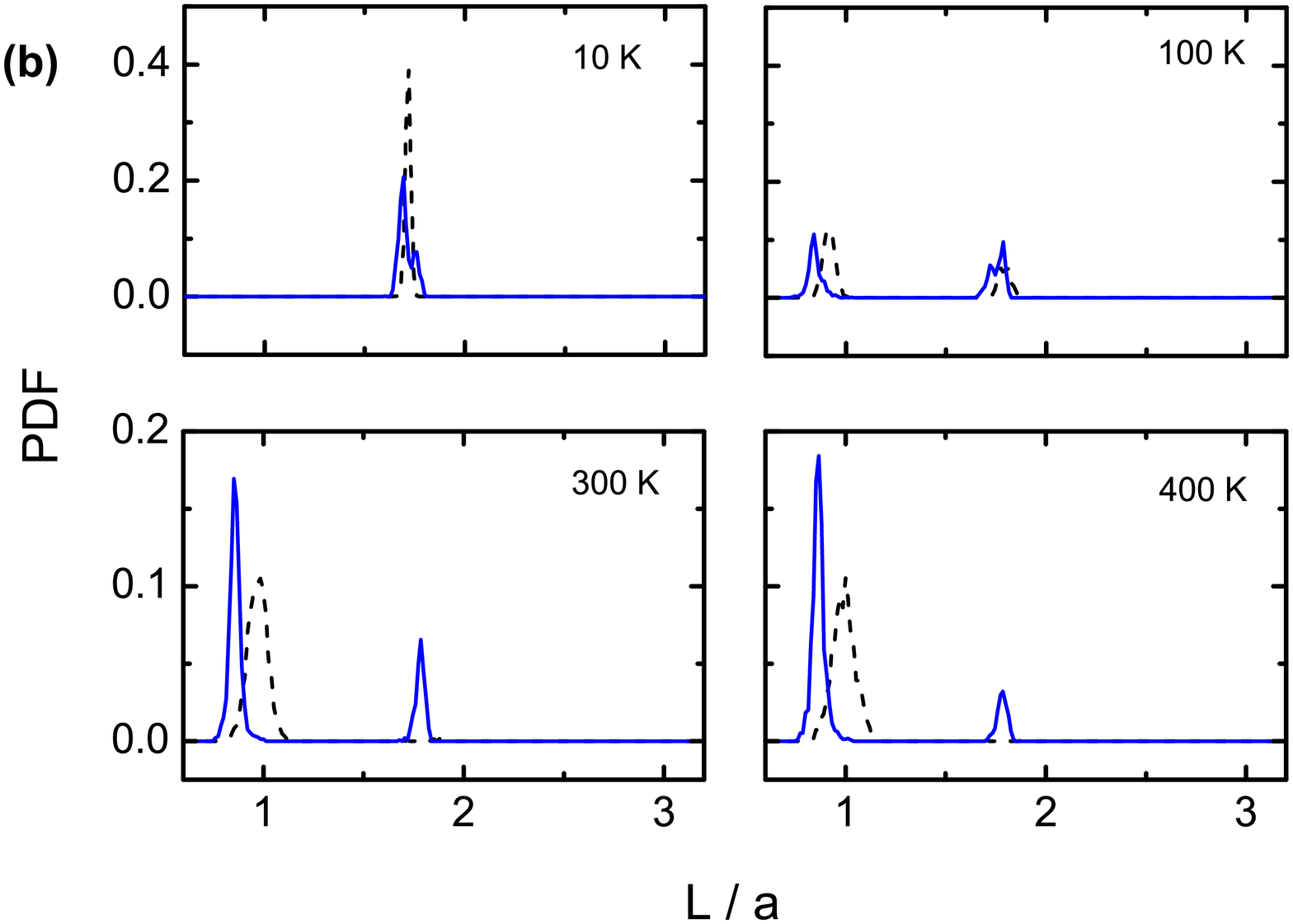}
\includegraphics[scale=0.25]{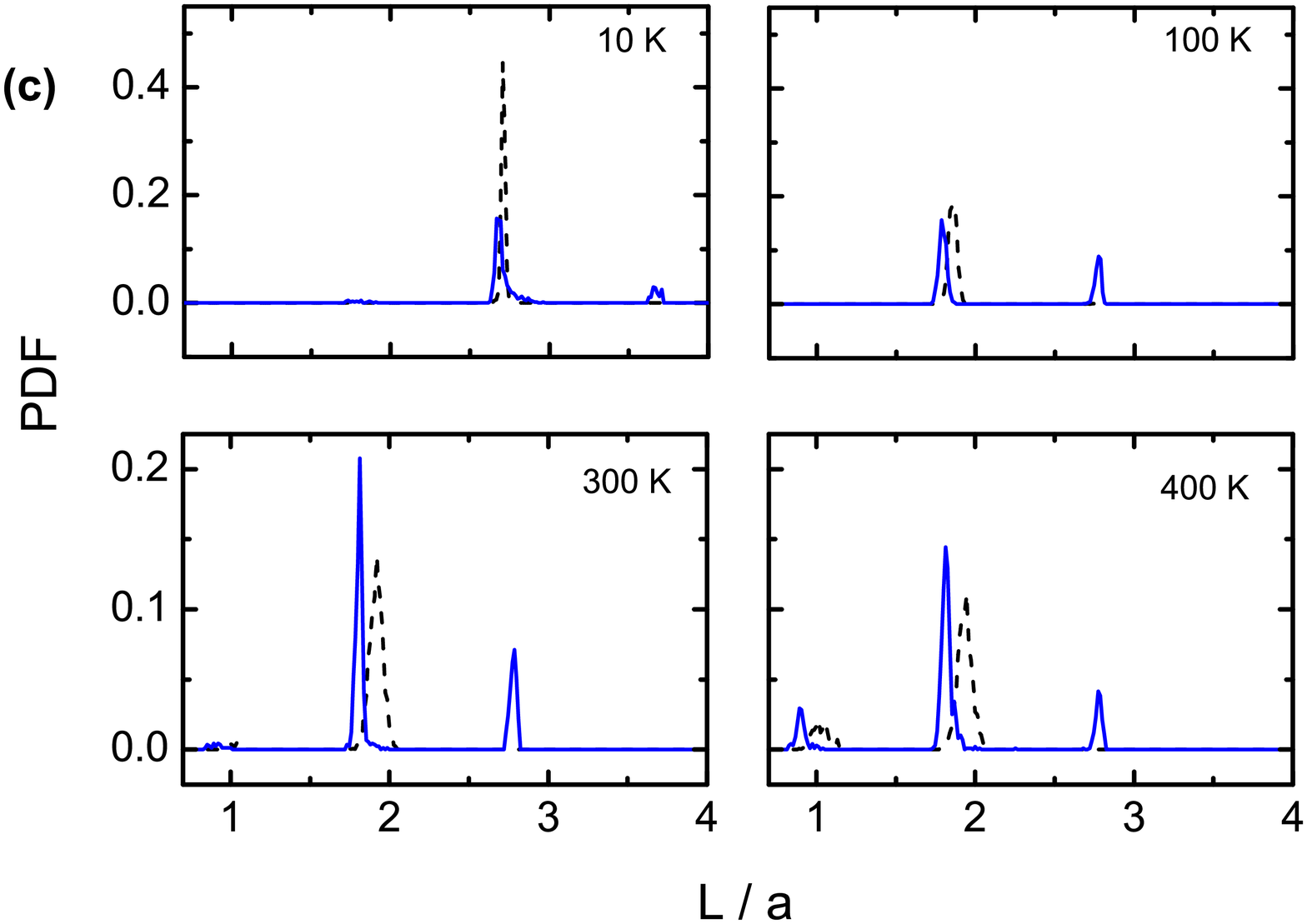}
\includegraphics[scale=0.25]{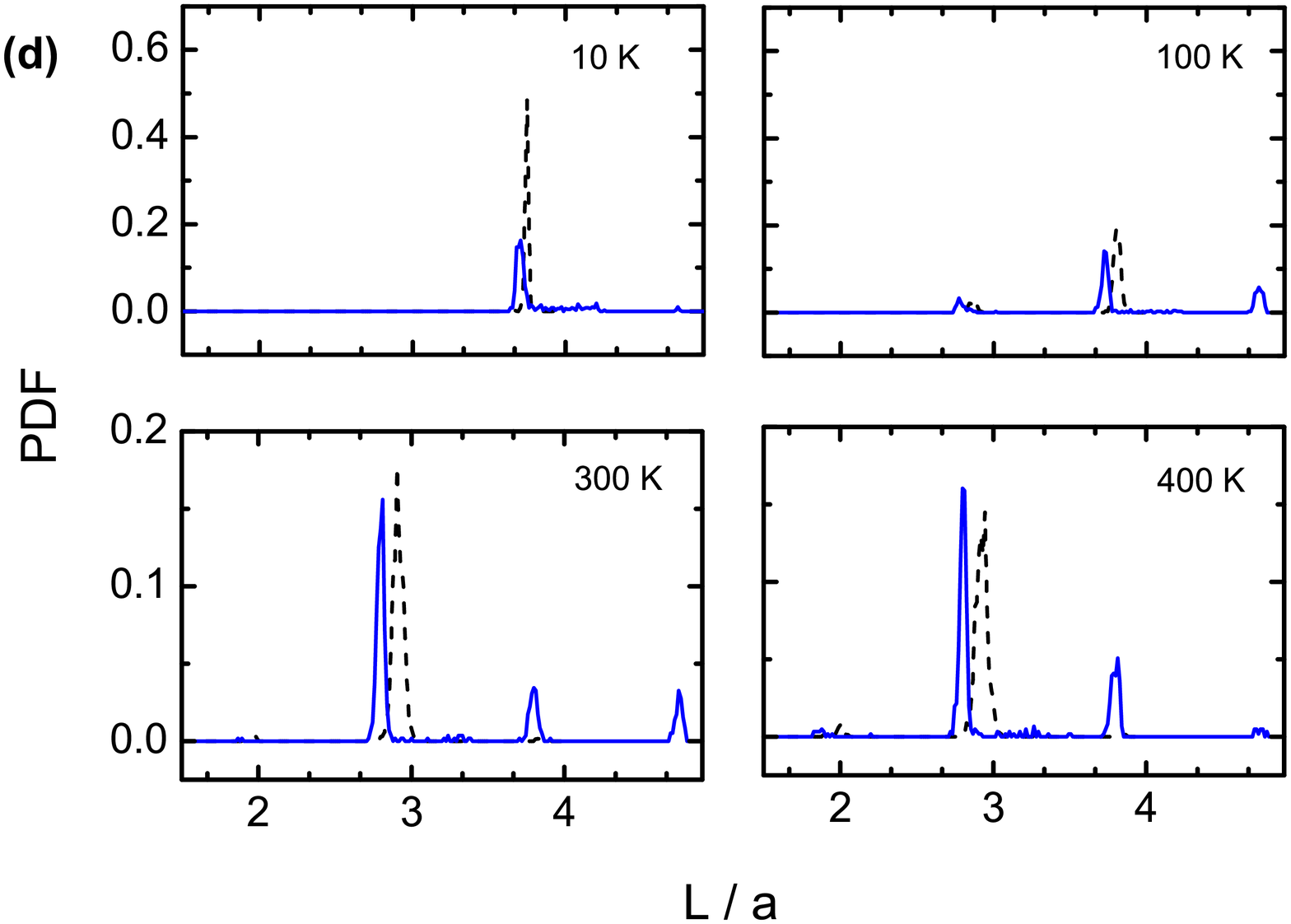}
\caption{ Slip length PDFs for the defect-free lattice
  (black dashed line) and the type I defect-in-minima one (blue line)
  at: (a) $U_0= 0.27$, (b) $U_0=0.38$, (c) $U_0= 0.58$, and (d)
  $U_0=0.88$ eV. For each potential value above we show the PDF at
  $T=10$, $100$, $300$ and $400$ K.}
\label{Fig8}
\end{figure*} 

\section{Surface defects with temperature}

In this section we will present results on the combined effect of
surface defects and temperature in the friction force of the
system. Theoretical and experimental results have shown that thermal
effects are fundamental to understand friction at this
scale~\cite{GneccoPRB2000,SangPRL2001,TshiprutPRL2009,SillPRL2003,SchirmeisenApplPhysLett2006,RiedoNanotech2004,OFajardoPRB2010}.
Previous studies show that the main temperature effect is a reduction
of the friction force in the stick-slip region. This reduction, which
has been observed at high temperatures, is explained in terms of
thermally activated jumps of the tip~\cite{GneccoPRB2000,SangPRL2001}.
Recently, it has been also observed that, at low temperatures, thermal
fluctuation can increase the friction force, which reaches a maximum
and then
decreases~\cite{OFajardoPRB2010,TshiprutPRL2009,SchirmeisenApplPhysLett2006}.
This effect is understood in terms of a reduction of the mean length
of the slip events which dominates at low temperatures.

\begin{figure*}[tb]
\centering
\includegraphics[scale=0.7]{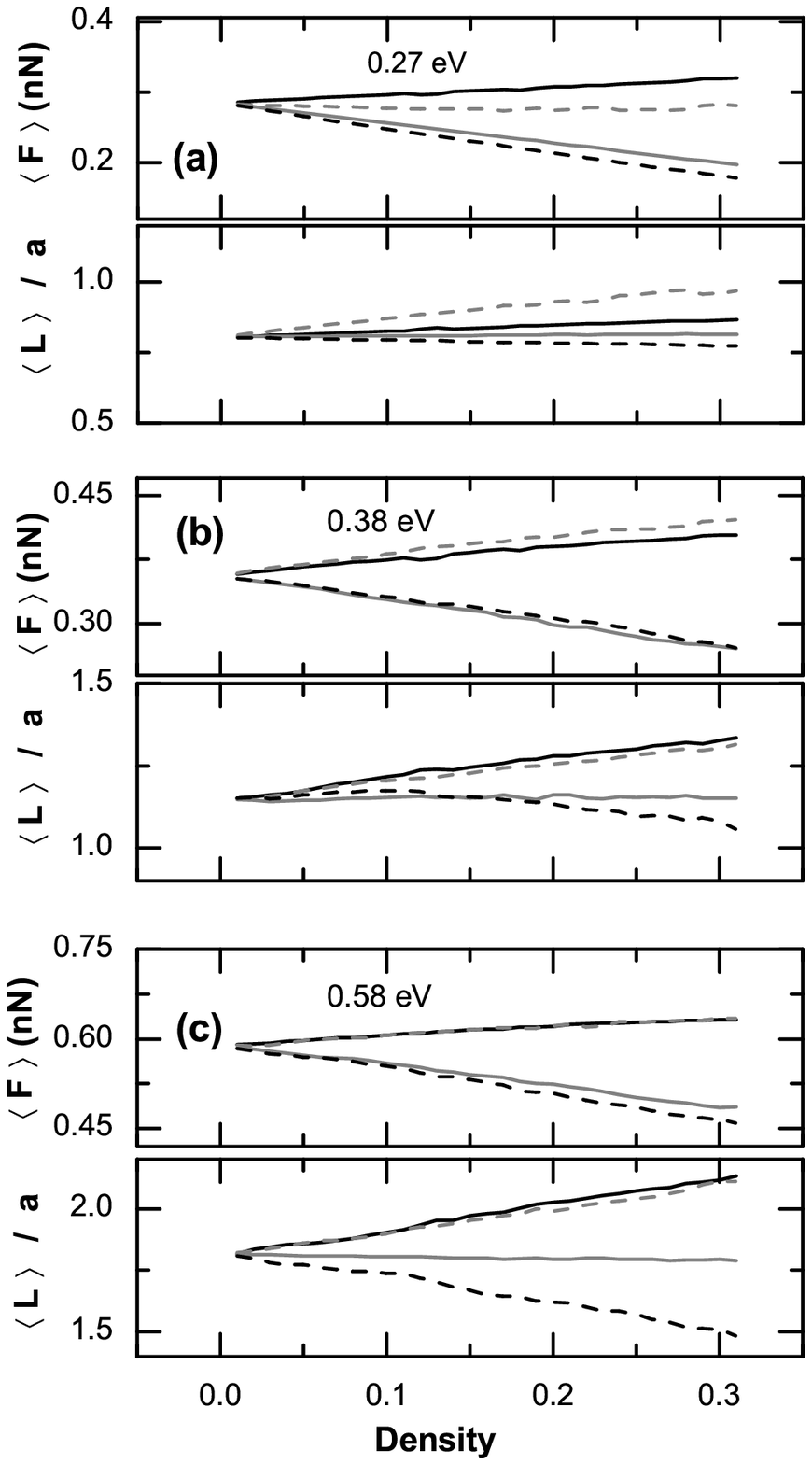}
\includegraphics[scale=0.7]{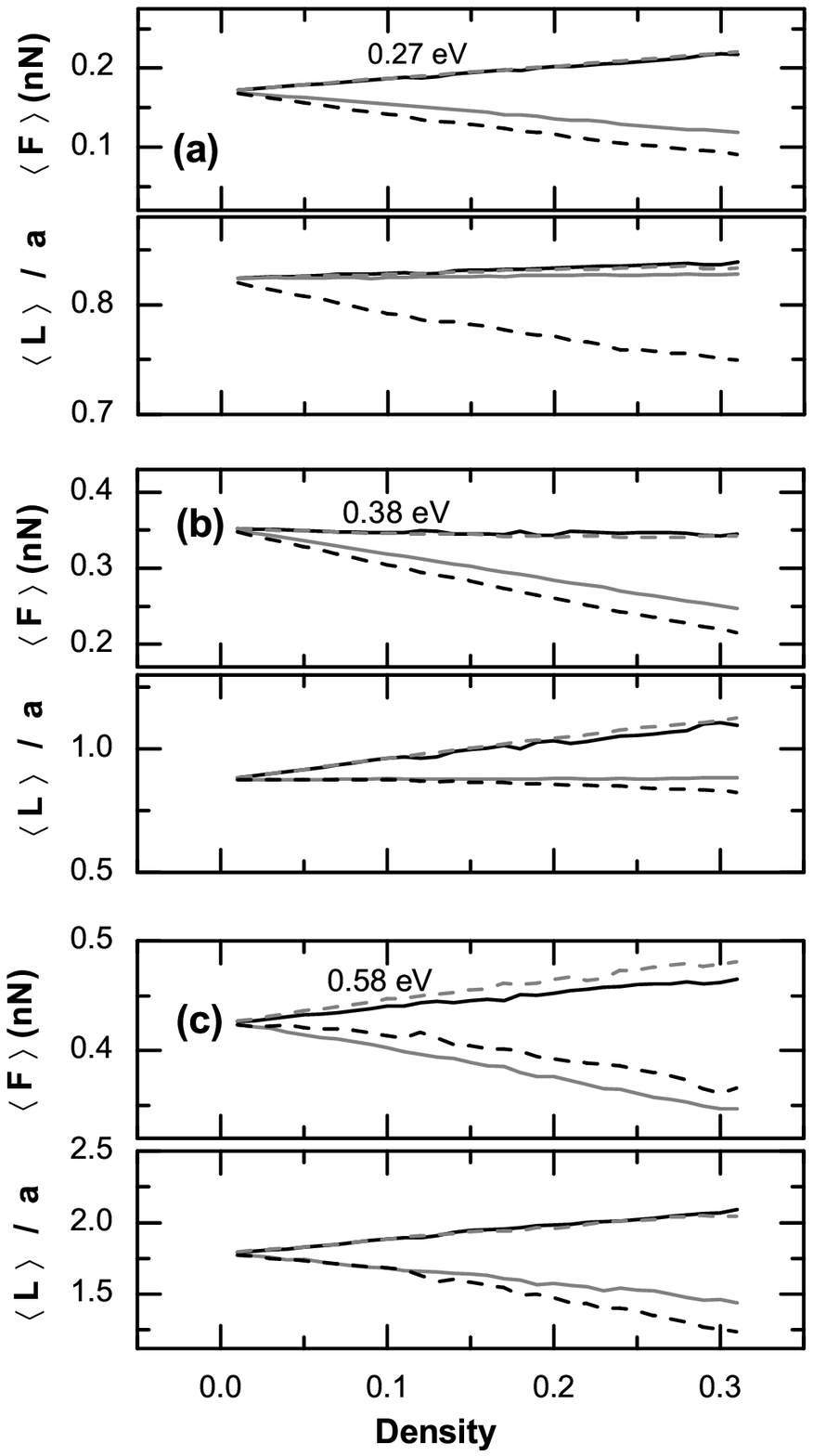}
\caption{ Average friction force and average length
  slip versus density of defects centered in minima (gray) or in
  maxima (black) . Type I defects are continuous lines and type II
  dashed ones. We show results for $ Uo= 0.27$ eV, $U_0=0.38$ eV,
  $U_0=0.58$ eV at $v_s=10$nm/s and $T=100$K (left) and 300K (right).}
\label{Fig9}
\end{figure*} 

Due to the important role played by temperature it is natural to
consider now the combined effect of thermal fluctuations and surface
defects in the response of the system. Figures~\ref{Fig6},~\ref{Fig7},~\ref{Fig8},
and~\ref{Fig9} summarize the main results.  In figure~\ref{Fig6}(a) we
compare at $T=0$, $50$ and $300$ K, the average friction force versus
the potential amplitude $U_0$ for the defect-free lattice (open
symbols) and the type I defect-in-minima. In figure~\ref{Fig6}(b) we
show the same comparison for the type I defect-in-maxima
case. Differences are clearly visible in all the studied temperature
range.  This effect is also studied in figure~\ref{Fig7} where we show
the average friction force $\langle F \rangle$ and slip length
$\langle L \rangle$ as a function of the system temperature for four
different values of $U_0$ corresponding to the four different
dynamical states marked in figure~\ref{Fig2}.

The inclusion of surface defects does not modify the two main
competing thermal effects already reported in the literature. The low
$U_0$ and high temperature regimens are dominated by a monotonic
decreasing of friction due to thermal activation without modification
of the mean slip length. At low temperatures and high $U_0$ values,
friction increases due to the reduction of the mean slip
length~\cite{OFajardoPRB2010}. However we observe that at a given
temperature the friction force for the type I defect-in-minima case is
bigger than for the defect-free one which in turns is also bigger than
for the type I defect-in-maxima lattice.
\begin{figure*}[tb]
\centering
\includegraphics[scale=0.26]{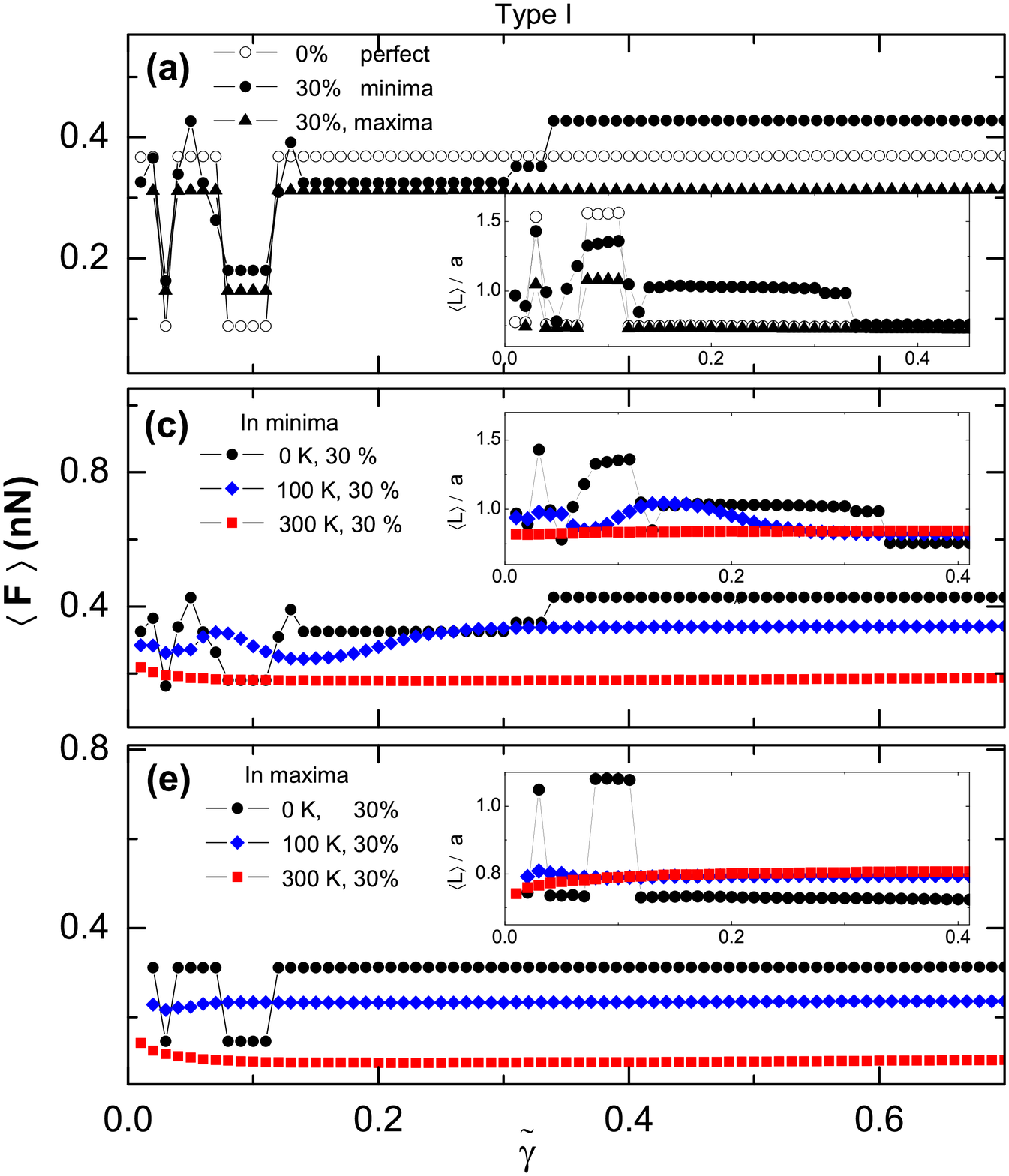}
\includegraphics[scale=0.26]{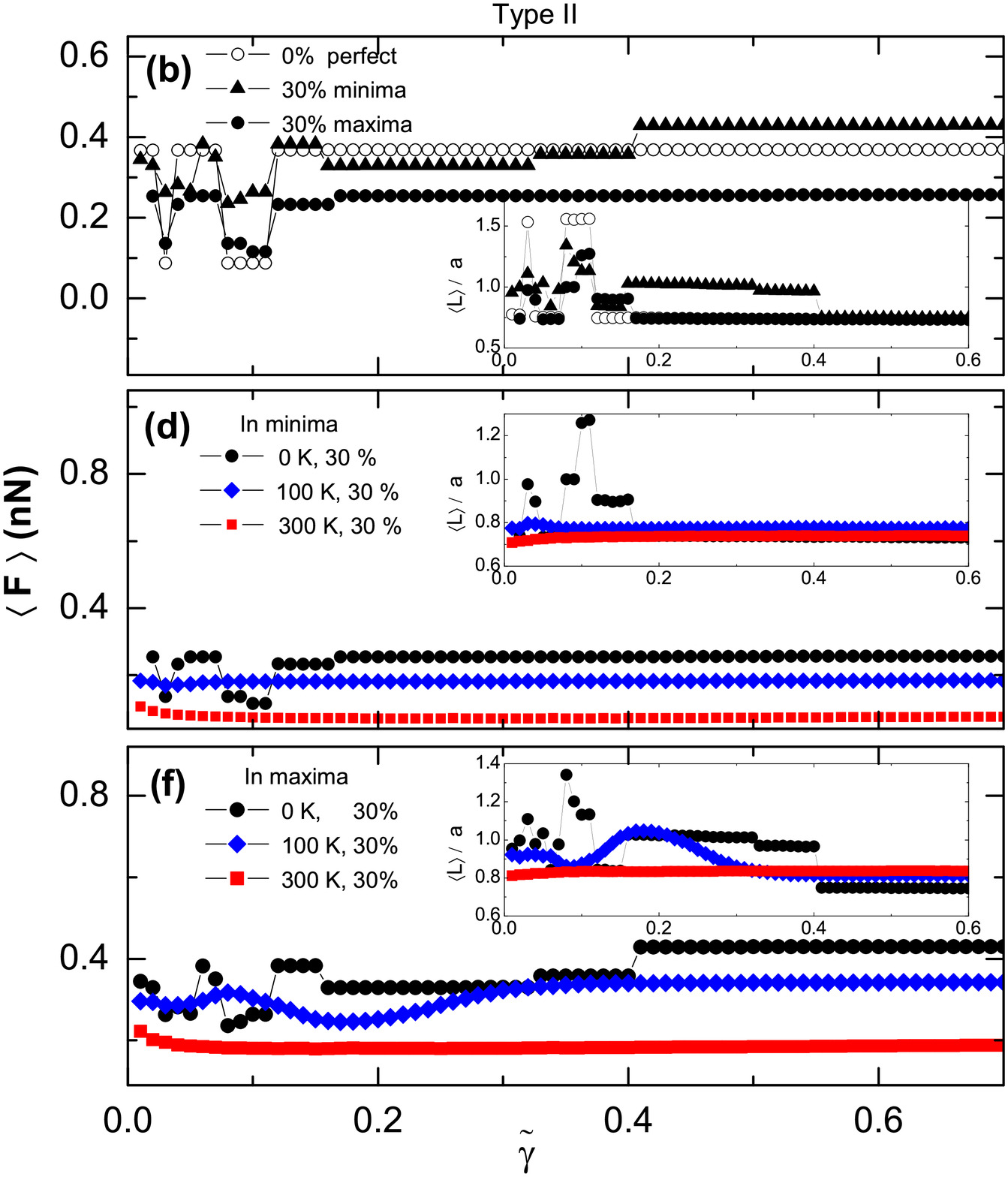}
\caption{ The mean friction force and mean slip length
  versus normalized damping at $T=0$ K for perfect lattice (open
  symbols), defects in minima of lattice and defects in maxima of
  lattice. (a) Type I defects and (b) type II defects.  Combined
  effect of surface defects and temperature (circles for $T=0$,
  diamonds for $100$ and squares for $300$ K) on the mean friction
  force and mean slip length as a function of normalized damping for
  defects in minima and defects in maxima of lattice. (c),(e) for type
  I and (d),(f) for type II defects. In all cases $U_0=0.27$ eV and
  $30$ \% for density of defects.}
\label{Fig10}
\end{figure*} 

In figure~\ref{Fig8} we show the change in the probability distribution
functions by the combined effect of defects and temperature. We
present results for type I defect-in-minima. The figure shows the
results for the same $U_0$ values used in figure~\ref{Fig5} but now at
four different temperature values. In the first case ($U_0=0.27$ eV),
figure~\ref{Fig8}(a), temperature suppress the second peak of the PDF. In
this case temperature affects more importantly the defect-free case
and moves apart the location of the peaks.  For $U_0=0.38$ eV,
figure~\ref{Fig8}(b), temperature activates a second peak close to
$\langle L \rangle=1$. As temperature increases the original one
decreases and disappears for the defect-free lattice case.  For
$U_0=0.58$ eV, figure~\ref{Fig8}(c), at zero temperature the main peak
is observed close to $\langle L \rangle=3$. At higher temperatures we
observe first the appearance of a second peak close to $\langle L
\rangle=2$, which will become the more important one, and later a
third one close to $\langle L \rangle=1$. At $U_0=0.88$ eV,
figure~\ref{Fig8}(d), the normalized damping is smaller and dynamics is
more complex and oscillations between adjacent wells are possible.

An important consequence of figures~\ref{Fig7} and ~\ref{Fig8} is that
although the PDF for the lattice with and without defects can be quite
similar (especially at low $T$) the mean friction force is very
different due to reduction of the effective barrier of the system.

Finally, we have studied again the dependence of the friction force
and mean slip length as a function of the density of defects in the
surface. In figure~\ref{Fig4} we showed the results for the zero
temperature case. Now, in figure~\ref{Fig9} we plot the similar results
for $T=100$K and $T=300$K. As we can see in the figures at all the
temperatures, both, the mean friction force and the mean slip length
depend almost linearly on the density of defects for the studied range

\section{Damping effect}

The effect of coupling of the system to many other degrees of freedom
is modeled by assuming the existence of a heat bath at a given
temperature and satisfying the fluctuation-dissipation relation. The
dissipation associated with this coupling is controlled by the damping
parameter, with temperature being a well-controlled parameter. However
to determine the correct value of the effective damping for the tip is
a much more complex and difficult question. Thus, it is important to
study the dynamics of the system for different values of the damping
in order to check the robustness of the results against changes of
this important parameter of the dynamics of the system.

A critical damping value is usually defined above which the tip
oscillations disappear (overdamped dynamics). In this case all the
deterministic slips have a length close to one lattice
constant. However, there are experimental evidences which indicate
that the damping of the system is below this critical value. In this
case the dynamics is more complex and
richer~\cite{KerssemakersApplPhysLett1995,RRothTribolLett2009}. We
have studied first in the regular case the effect of the damping
parameter on the friction force for values of $\widetilde{\gamma}$
crossing the critical value. Then we analyze how this behavior is
affected when surface defects are included. In figure~\ref{Fig10}(a) we
observe a series of steps in the friction force versus damping curve
(open symbols). Close to the discontinuities a small change in the
damping can produce a large change in the friction force. Every step
in figure corresponds to a slip close to one or two lattice sites
respectively.  In the overdamped limit only the first available
minimum (site 1) is reached. When smaller values of the damping are
achieved the tip can reach a further minimum (site 2). If damping is
reduced again the tip can move through 1 to 2 and then oscillates back
to 1 where is trapped. At small damping the tip oscillates back and
forth between both minima before reaching an equilibrium value. Then,
by decreasing the value of the damping more transitions are observed.
For other values of the parameters (larger $U_0$), more complex
situations are found. In these cases dynamics results from the
interplay between a larger number of minima than that shown in
figure~\ref{Fig10}.

When surface defects are included (full symbols), additional slight
modifications are observed. Figure~\ref{Fig10}(a) shows the change of
the friction force and the mean slip length as a function of the
damping for a density of $30$\% of defects and it compares to the
regular case curve. With inclusion of the thermal effects (diamond
symbols, $100$ K), the steps are strongly smoothed, see
figure~\ref{Fig10}(b). At higher temperatures steps are strongly smoothed
or even suppressed (square symbols, $300$ K).

\section{Discussion and conclusions}

We have studied the effect of the presence of four different types of
surface defects on atomic friction. Our results show that significant
changes can be observed even at high temperature. As expected,
differences are more important for $U_0$ values close to deterministic
dynamical transition points of the system. 
\begin{figure*}[tb]
\centering
\includegraphics[scale=0.21]{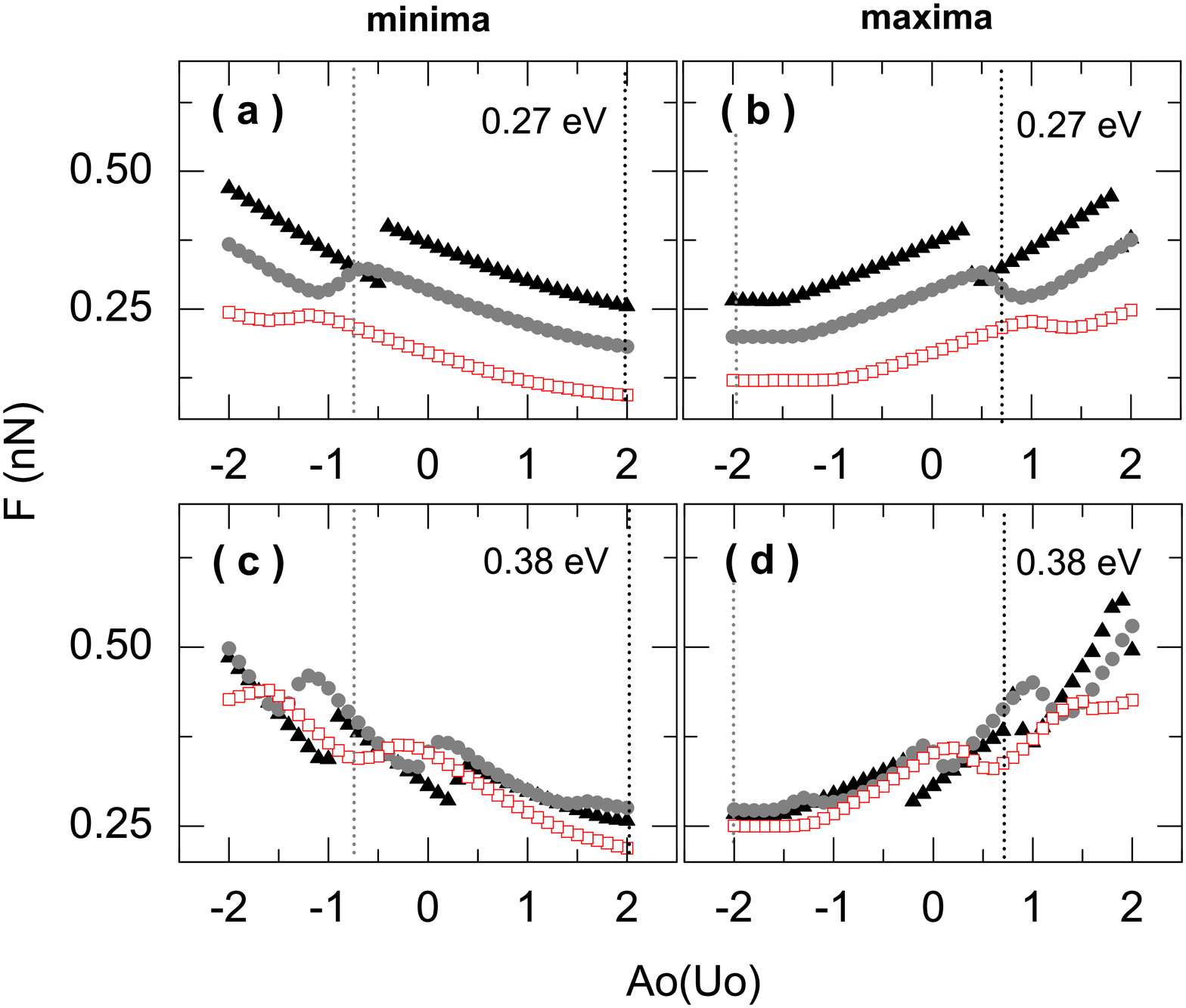}
\includegraphics[scale=0.21]{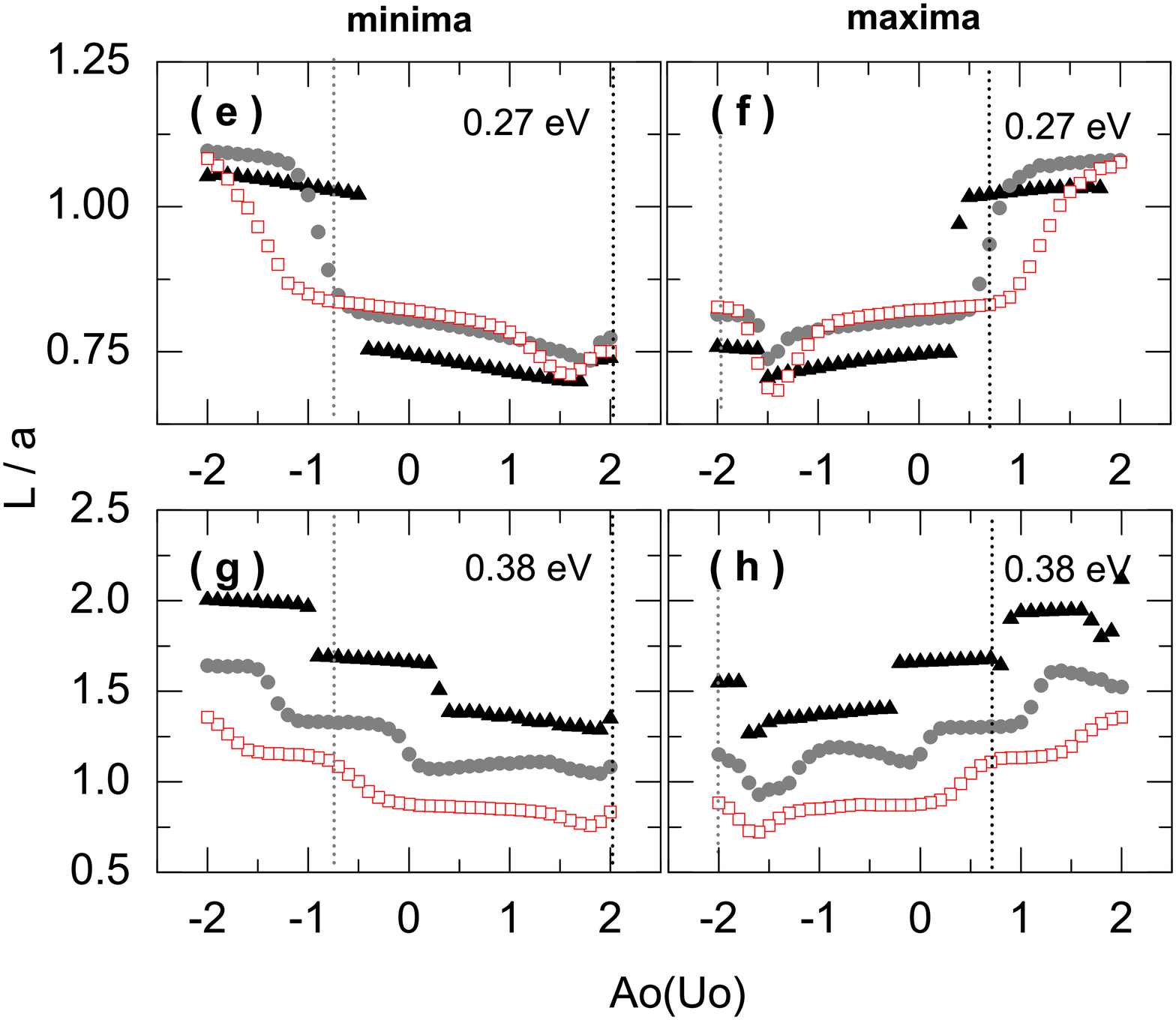}
\caption{ Friction force and mean slip length dependence
  on the amplitude of the defect potential $A_0$ for defects in minima
  and in maxima at $T=0$ (black triangles), 100 (grey circles) and 300K
  (red squares)}
\label{Fig11}
\end{figure*}

As we have seen, for surface defects, the change in the mean friction
force can not be understood in terms of the change in the mean slip
length.  In order to understand our numerical results we have done an
study of the probability distribution function of the slip length. A
rich scenario where jumps of very different length may coexist is
found.

We have identified two main different mechanisms which modify the mean
friction force of the system when defects are considered. First,
defects modify locally the potential profile in a way that change
importantly the instantaneous friction force that the tip experiences
when crosses one defect. Second, the presence of defect also changes
the slip length probability distribution which also changes the mean
friction force.

With respect to the density of defects in the system, our results show
an almost linear dependence on this parameter. We will discuss now on
a different point which is related with the amplitude of the defect
potential given by $A_0$. We have chosen $A_0=-0.71U_0$ for type I
defect in minima and $A_0= 0.71U_0$ for type II defect in maxima and
$A_0=-2U_0$ for type I defect in maxima and $A_0= 2U_0$ for type II
defect in minima. Those values were chosen to get the potential
profiles showed in figure~\ref{Fig1}. However other values of $A_0$ can
be of interest for a particular real situation. Figure~\ref{Fig11}
shows the results of the mean friction force and mean slip length
dependence on the defect amplitude coefficient $A_0$. Type I defects
correspond to $A_0<0$ values and type II to $A_0>0$ values. Results
are given for three temperature values (0, 100 and 300K) and two $U_0$
values (0.27eV and 0.38ev), showing two different physical
situations. The figure shows that the friction force also depends
importantly on this parameter in a way that is comparable to the
effect of temperature or density defects. Figure also shows that, as
expected, results for defect in minima and in maxima with different
sign of $A_0$ are very similar. This was also shown in
figures~\ref{Fig4}, \ref{Fig7} and \ref{Fig9} for instance.

\begin{figure*}[tb]
\centering
\includegraphics[scale=0.35]{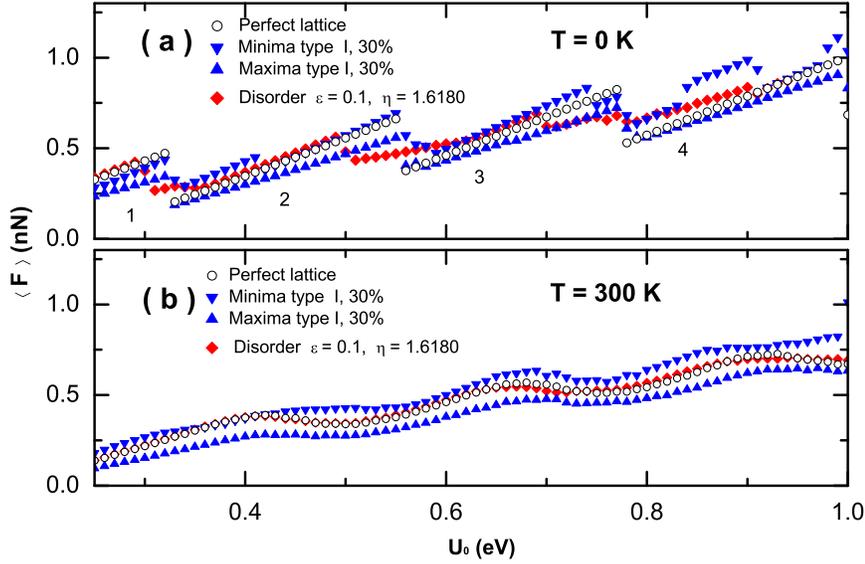}
\caption{ (a) Mean friction force versus corrugation
  potential amplitude $U_0$ at $T=0$ K. We compare the effects from
  two representative types of surface defects on the friction
  force-corrugation potential curve of the perfect lattice (open
  circles): type I surface defects in maxima (solid triangles up), in
  minima (solid triangles down) and the effect of surface disorder (solid
  diamonds). In (b) the similar comparison at $T=300$ K.}
\label{Fig12}
\end{figure*}

To finish, we find interesting to compare our results for surface
defects to the recently studied effect of surface
disorder~\cite{OFajardoPRB2010}. There, surface disorder was modeled
including a small second harmonic term in the standard tip{-}surface
interaction potential,
\begin{equation}
 V(x)=-U_o \left[ 1.0 + \epsilon\ \sin\left( \frac{2\pi x}{b}\right)
   \right] \cos\left( \frac{2\pi x}{a}\right).
\label{disorder}
\end{equation} 	
Thus the surface disorder changes the potential profile, introduces a
distribution of barrier heights and moves slightly the positions of
the potential maxima and minima. 

Figure~\ref{Fig12} compares the results obtained for the case of the
perfect lattice, surface disorder and surface defects. It shows the
results for $T=0$K y $T=300$K and at different values of $U_0$. The
figure shows results for type I defects in minima and maxima. Similar
results for other types of defects are found.  At low temperature we
found that it is not easy to distinguish between the presence of
defects and disorder.  However, an analysis of instantaneous
quantities such as force and tip position allows to discriminate both
situations. At high enough temperature our results indicate that the
role of the surface disorder is screened by the thermal effects but
the presence of surface defects modifies the friction force of the
system. In the regular and in the lattice with disorder the mean value
of the barrier that the tip experiences is the same. However the
presence of surface defects increases or decreases, depending on the
type of the defects, the value of such mean barrier.

Our model accounts for the effect of adsorbed molecules or vacancies
in the lattice, but it could be used also to study artificially
designed grooves or ridges. Our results show that a significant
density of defects in a sample may change the friction force curve at
any temperature and thus such presence affects a comparison between
numerical and experimental results.

However, it is important to point that friction is a very general phenomenon which occurs in several
problems. Because of that, the study of simple models is important
since it allows for understanding basic effects that can be relevant
or realized in different kind of systems. This is the case for
instance of the motion of a single colloidal particle in a colloidal
system. There are now experiments and simulations where a single
colloidal particle can be driven through other particles or quenched
disorder and the effective friction and drag measured for varied
temperature, density, pull rate and so forth~\cite{HabdasEurophys04,ReichhardtPRL04,ReichhardtPRE08,MejiaSoft11}. 
Some similar variations on the system can be rapidly realized 
by driving a single particle over a random, periodic or 
combination of periodic and random substrates in 1D, 2D and even 3D 
so that some of the results from the present work 
could be tested more directly on real systems.

Another system where the results of the present paper could be applied
is that of vortices in type-II superconductors. Very recent
experiments drag individual vortices over random or periodic
pinning arrays~\cite{HoffmanAPL08,AuslaenderNature09,ReichhardtNature09}. It would be interesting to make a periodic 1D or 2D
pinning potential and drag a single vortex. This experiment could then
be repeated for adding additional defects that could act like
attractive or repulsive sites.

\ack

The authors acknowledge financial support from Spanish MICINN through project
No. FIS2008-01240, cofinanced by FEDER funds. O. Y. Fajardo acknowledges financial 
support from FPU grant by Spanish MICINN. 

\section*{References}


\begin{thebibliography}{34}
\bibitem{Comentario2}
A change in $U_o$ also changes the effective damping of the system since $
  \widetilde{\gamma} = \gamma / \omega_p$ and $\omega_p=2\pi \sqrt{U_0/Ma^2}$.
  Thus a high load implies a smaller effective damping and it allows longer
  slip events
\bibitem{Comentario3}
We want to mention that due to the existence of an early creep stage in the
  movement of the tip before each slip event, the slip length is not exactly
  equal to a multiple of the lattice constant
\end{thebibliography}

\end{document}